\documentstyle[PASJadd,psfig]{PASJ95}
\newcommand{\vol}{}
\newcommand{\no}{}
\newcommand{\lett}{}
\newcommand{\spage}{}
\newcommand{\rdate}{1999 August 18}
\newcommand{\adate}{1999 September 27}
\begin{document}
\title{Discovery of an X-Ray Pulsar in the SMC: AX J0058$-$7203}
\author{Masahiro {\sc Tsujimoto}, Kensuke {\sc Imanishi}, Jun {\sc
Yokogawa}, and Katsuji {\sc Koyama}\thanks{CREST, Japan Science and
Technology Corporation (JST), 4-1-8 Honmachi, Kawaguchi, Saitama, 332-0012}\\
{\it Department of Physics, Graduate School of Science, Kyoto
University, Sakyo-ku, Kyoto, 606-8502} \\
{\it E-mail(MT): tsujimot@cr.scphys.kyoto-u.ac.jp}}
\abst{We report on the discovery and analysis of an X-ray pulsar, 
AX~J0058$-$7203, in the Small Magellanic Cloud. This pulsar exhibits
coherent pulsations at $P= 280.3$~s with a double-peak
structure. The X-ray spectrum is well fitted with a simple power-law
model of photon index $\Gamma \sim 0.7$.  No significant change of
the pulsation period over the observation was found. A comparison with
ROSAT observations in the same field reveals  that
AX~J0058$-$7203 is highly variable, and is most likely a Be star
binary pulsar.}
\kword{Magellanic Clouds --- Pulsars: individual (AX~J0058$-$7203) 
--- X-rays: binaries}
\maketitle
\thispagestyle{headings}

\section{Introduction}
X-ray binary pulsars (here XBPs) constitute a bright class of X-ray
sources in the sky, and have been major objects for X-ray
astronomy.  Sporadic mass accretion from a companion star onto a
spinning neutron star makes this class to be variable X-ray
sources. X-ray binaries with a Be star companion (here Be-XBPs) are
the most variable subclass; occasional outbursts may be caused by
mass-ejection episodes of a companion Be star or by 
encounters with a dense stellar wind region along with a highly
eccentric orbit of a neutron star. The average luminosity is
generally lower than ordinal XBPs, and is typically in the range of
$10^{34}$--$10^{35}$ erg~s$^{-1}$. This moderate luminosity
together with a limited duty ratio of the outbursts has prevented us to
perform a complete survey of this class with the conventional
non-imaging instruments. The imaging instruments on-board the
Einstein and ROSAT satellites greatly improved the detection
threshold. Since XBPs (also Be-XBPs) generally  exhibit a hard X-ray
spectrum,  contemporary hard X-ray satellites, ASCA (Advanced
Satellite for Cosmology and Astrophysics), RXTE (Rossi X-ray Timing
Explorer), and Beppo-SAX, further enhance the  detection probability
for this class of X-ray sources (XBPs and Be-XBPs).

In fact, multiple observations of the Small Magellanic Cloud (SMC)
caused a rush of X-ray pulsar discoveries (Yokogawa et al.\ 1999).
Among the known 16 X-ray pulsars in the SMC at present (Bildsten et
al.\ 1997; Kahabka et al.\ 1999; Lamb et al.\ 1999; Macomb et al.\ 
1999), 13 have been discovered  within 1.5  years: from  the end of
1997 to the middle of 1999.  Most are considered to be Be-XBPs based on 
their transient nature and/or an optical Be counterpart. 

We report here on the discovery and analysis of AX~J0058$-$7203, one of 
the newly discovered pulsars during the pulsar rush episode
(Yokogawa, Koyama\ 1998). We also examine the flux variability using
the archives of ASCA and ROSAT, and propose to classify this source
as a Be star binary system in the SMC.

\section{Observations}
ASCA observed a SMC region centered at R.A.$=00^{\rm h}59^{\rm
m}26^{\rm s}\hspace{-5pt}.\hspace{2pt}3$,
Dec.$=-72^{\circ}10^{\prime}12^{\prime\prime}\hspace{-5pt}.\hspace{2pt}5$ 
(equinox 2000) on 1997 November 14 to 15, during the AO-6 cycle. The
primary target of this observation was N66, one of SNRs in the
SMC region. Details concerning the ASCA satellite as well as its instruments (GIS:
Gas Imaging Spectrometer, SIS: Solid-state Imaging Spectrometer) and
X-ray telescope (XRT) can be found in Tanaka et al.\ (1994), Ohashi
et al.\ (1996), Makishima et al.\ (1996), Burke et al.\ (1991) and
Serlemitsos et al.\ (1995).  As is the nominal case, two  GISs  (GIS 2,
3) and SISs (SIS 0, 1) were operated in parallel.

AX~J0058$-$7203 was in the field of view of the GISs and SISs at
$\sim 13^{\prime}$ off-axis positions. We excluded the data taken
during the South Atlantic Anomaly or at an elevation angle less than
$5^{\circ} $  or the passage through a region where cut-off
rigidity is lower than 6 GeV. In addition, we employed a rise-time
discrimination technique to reduce particle events for the GIS
data. After these filterings, the net exposure times were  $\sim$ 35
ks for GIS and $\sim$ 31 ks for SIS.

\section{Results and Analysis}

\subsection{X-Ray Image and Source Identification}
A combined GIS image (GIS 2 + GIS 3) is shown in figure 1.  The
accurate position of this source was determined using two SIS
images (SIS 0 and 1) to be R.A.$=00^{\rm h}57^{\rm m}53^{\rm
s}\hspace{-5pt}.\hspace{2pt}3$,
Dec.$=-72^{\circ}02^{\prime}46^{\prime\prime}\hspace{-5pt}.\hspace{2pt}5$
with an error circle of $\sim 40^{\prime\prime}$ in radius
(Gotthelf 1996; ASCA News 4, 31), hence, 
we designate this source as AX~J0058$-$7203.
Two other X-ray pulsars are serendipitously located in the GIS field: 
1SAX~J0054.9$-$7226 with a pulsation periodof  58.969~s (Santangelo
et al.\ 1998) at the top left and 1SAX~J0103.2$-$7209 with a pulsation
period of 345.2 s (Israel et al.\ 1998) at the bottom right of figure
1. This demonstrates an  extremely high density of X-ray pulsars of
the  central region of the SMC.

%
%
\begin{figure}
\psfig{file=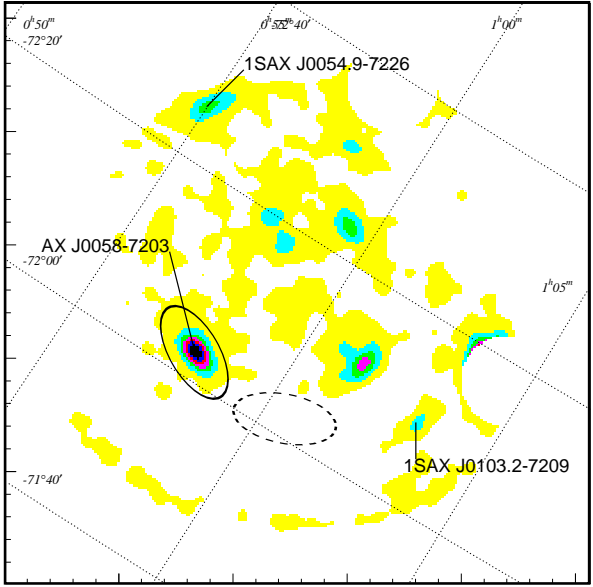,width=0.9\columnwidth}
\caption{GIS hard-band image (2.0--7.0 keV): X-ray photons from the
 source are extracted from the region shown with the solid line.
 The background spectrum was taken from the dotted region, where the
 off-axis angle is almost the same as that of the source
 region.  The calibration source regions of GIS 2 (at the top right)
 and GIS 3 (bottom right) are hollowed out.}
\end{figure}

For source identification, we checked two comprehensive SMC
source catalogs provided by Wang and Wu (1992) with  Einstein observatory,  
and by Kahabka et al.\ (1999) with ROSAT observatory.  
Source No.\,41 (SMC~0056.2$-$7219) in the former catalog and source No.\,124 
(RX~J0057.8$-$7202) in the latter one are found to be located within the error
circle of AX~J0058$-$7203.

\subsection{Temporal Analysis}
The X-ray photons of AX~J0058$-$7203 were extracted from the
elliptical regions, as is indicated by the solid line in figure 1 for
both the GIS and SIS data.  We first confirmed that there was no
flaring event nor any trend of flux variation during the observation.
After a barycentric correction of the photon-arrival times, we conducted
a Fast Fourier Transform (FFT), in order to search for any coherent
periodicity. Figure 2 shows the power spectrum for the GIS data  in
the 0.7--7.0 keV band. We found two significant peaks at
$3.6\times 10^{-3}$ Hz and $7.2 \times10^{-3}$ Hz. The chance
probabilities to obtain these two peaks  are estimated to be  $9 \times
10^{-10}$  and  $1 \times 10^{-11}$, respectively. Since the
frequency ratio between the two peaks is exactly double, we
infer  the lower frequency peak to be  the fundamental period of
$P= 280.3$~s and the higher to be the second harmonics.

%
%
\begin{figure}
\psfig{file=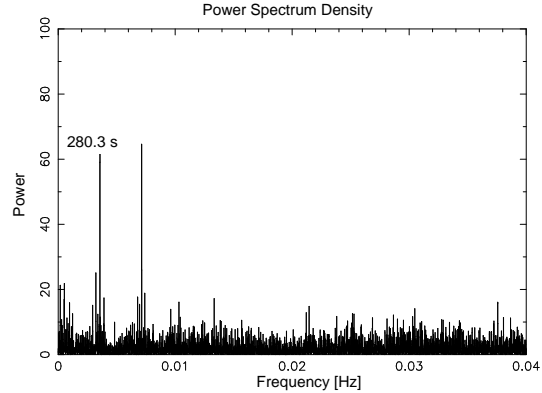,width=0.9\columnwidth}
\caption{Power Spectrum Density of AX~J0058$-$7203 for the GIS and SIS data
 in the 0.7--7.0 keV band.  The fundamental pulsation together
 with the secondary harmonic can be clearly seen.}
\end{figure}

We then employed a folding technique to the GIS and SIS data
around  a trial period of $P= 280.3$~s, and found the most likely
period to  be $280.4 \pm 0.3$ s. The folded pulse profile of
280.4 s period is given in figure 3.  In the pulse profile, we can see
two peaks, as was already expected from the power spectrum.  Since the
intensity of the main peak is about twice larger than that of the
sub-peak, we conclude that 280.4 s is really the fundamental pulse
period. The two-peak pulse profile does not change with the X-ray
energy. The pulse fraction, defined as $(I_{\rm max}-I_{\rm
min})/(I_{\rm max}+I_{\rm Imin})$, where $I_{\rm max}$ and $I_{\rm
min}$ are the maximum and the minimum count rate including
background photons, is 61\% in the soft band (0.7--2.0 keV) and
48\% in the hard band (2.0--7.0 keV).
We also divided the full observation data into two data sets, the
former half and the latter half, then separately applied the folding
technic, but found no significant changes of the pulsation period
over the observation period.
%

%
%
\begin{figure}
\psfig{file=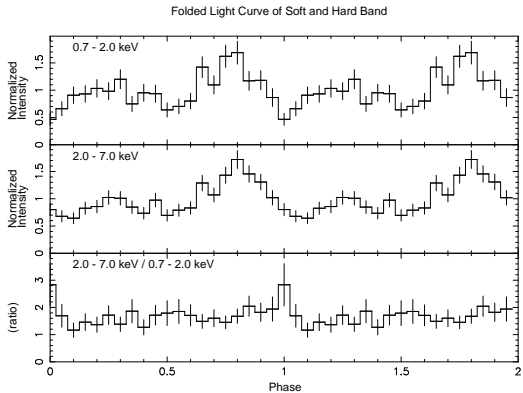,width=0.9\columnwidth}
\caption{Pulse profiles in the two cycles for the soft band (0.7--2.0
 keV: top panel), the hard band (2.0--7.0 keV: middle panel), and
 the intensity ratio of the  two bands (bottom panel). The pulse
 profiles were constructed by using both the GIS and SIS data.}
\end{figure}

\subsection{Spectral Analysis}
Using the same source data as in the timing analysis (solid line in
figure 1) and subtracting the background  data  in the dotted-line
region in figure 1, we constructed the GIS and SIS X-ray spectra
separately, showing the result in figure 4 with filled squares and
circles, respectively.  For our spectral analysis, both the GIS and
SIS data in the energy band from 1.0 keV to 8.0 keV were binned so
that each bin has at least 40 photons, and were fitted
simultaneously. 

For the continuum emission, we applied a power-law model with
interstellar absorption, which is the ``standard'' model for X-ray
binary pulsars (Nagase\ 1989), and obtained an acceptable fit.  The
best-fit parameters are given in table 1.

%
%

In figure 4, we can see a hint of excess at the energy of the K$\alpha$
line of He-like aluminum. To investigate the existence of this line 
emission, we fitted the spectrum with the absorbed power law with a
Gaussian line centered at 1.59 keV (K$\alpha$ line of He-like
aluminum). However, the significance level of this emission is less
than 1-$\sigma$ level. We also fitted the spectrum with the line-center 
variable, but no significant emission line (more than
1-$\sigma$ level) was found.

The X-ray flux in the  0.7--10.0 keV band  is calculated to be
$\sim 3.0 \times 10^{-12}$ erg~cm$^{-2}$~s$^{-1}$ based on the best-fit
parameters, which indicates an absorption-corrected luminosity of
$\sim 1.4 \times 10^{36}$ erg~s$^{-1}$, assuming the distance to
the SMC to be 60 kpc (Mathewson\ 1985).

%
%
\begin{figure}
\psfig{file=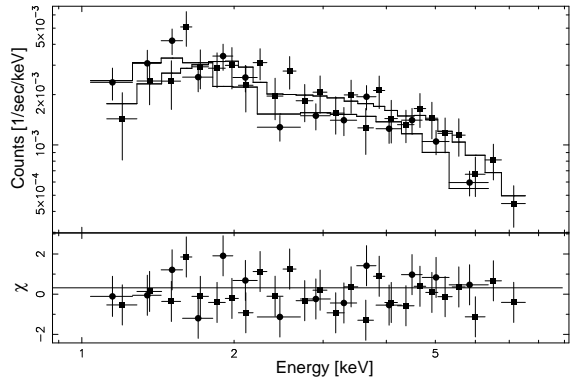,width=0.9\columnwidth}
\caption{Upper panel: The GIS (filled squares) and SIS (filled circles)
 spectra  of AX~J0058$-$7203.  The best-fit power law models of the
 simultaneous fitting are indicated by the solid lines. Lower panel:
 Residuals from the best-fit models.}
\end{figure}

\section{Discussion}
A long pulse period and a rather flat spectrum indicate this source to
be an accretion-powered binary pulsar with a high-mass companion
(Nagase 1989). Among high-mass binary pulsars with various types of
optical companion, binary systems of an X-ray pulsar with a
Be star share the transient nature which is exclusively
found in this subclass (Stella et al.\ 1986). 

Thus, information concerning the long-term flux variability of this source is
critical to reveal its nature. We therefore used ROSAT and Einstein
archival data which cover the region of AX~J0058$-$7203 to search
for its history of variability.  The count rate, or upper limit, from
this source are summarized in table 2 together with the ASCA result.

%
%

>From table 2, we find that AX~J0058$-$7203 shows a variation of
more than 10 times in flux, which favors AX~J0058$-$7203
to be a Be star binary pulsar.  Due to limited data from this source 
 the Be star interpretation still remains preliminary; hence, follow-up
observations invollving long-term pulsation monitoring and optical
identification of a companion star are encouraged.

\vspace{1pc}\par
We thank all members of the ASCA team. 
This research has made use of data obtained through the High Energy
Astrophysics Science Archive Research Center Online Service,
provided by the NASA/Goddard Space Flight Center. J.Y.\  is
supported by JSPS Research Fellowship for Young Scientists.

\clearpage
\section*{References} 
\re 
Bildsten L., Chakrabarty D., Chiu J., Finger M.H., Koh D.T., Nelson
R.W., Prince T.A., Rubin B.C. et al.\ 1997, ApJS\ 113, 367
\re			       
Burke B.E., Mountain R.W., Harrison D.C., Bautz M.W., Doly J.P.,
Ricker G.R., Danniels P.J.\ 1991, IEEE Trans.\ ED-38, 1069			       
\re
Israel G.L., Stella L., Campana S., Covino S., Ricci D., Oosterbroek
T.\ 1998, IAU Circ.\ 6999
\re			       
Kahabka P., Pietsch W., Filipovi\'{c} M.D., Haberl F.\ 1999, A\&AS\ 136, 81
\re
Lamb R.C., Prince T.A., Macomb D.J., Finger M.H.\ 1999, IAU Circ. 7081
\re
Macomb D.J., Finger M.H., Harmon B.A., Lamb R.C., Prince T.A.\ 1999, ApJ 518, L99
\re
Makishima K., Tashiro M., Ebisawa K., Ezawa H., Fukazawa Y., Gunji
S., Hirayama M., Idesawa E.\ 1996, PASJ\ 48, 171
\re			       
Mathewson D.S.\ 1985, Proc.\ Astron.\ Soc.\ Aust.\ 6, 104
\re
Nagase F.\ 1989, PASJ\ 41, 1
\re
Ohashi T., Ebisawa K., Fukazawa Y., Hiyoshi K., Horii M., Ikebe Y.,
Ikeda, H., Inoue H. et al.\ 1996, PASJ\ 48, 157 
\re
Santangelo A., Cusumano G., Israel G.L., Fiume D.D., Orlandini M.,
Frontera F., Parmar A.N., Marshall F.E. et al.\ 1998, IAU Circ.\ 6818 
\re
Serlemitsos P.J., Jalota L., Soong Y., Kunieda H., Tawara Y.,
Tsusaka Y., Suzuki H., Sakima Y. et al.\ 1995, PASJ\ 47, 105
\re
Stella L., White N.E., Rosner R.\ 1986, ApJ\ 308, 669
\re
Tanaka Y., Inoue H., Holt S.S.\ 1994, PASJ\ 46, L37
\re			       
Wang Q., Wu X.\ 1992, ApJS\ 78, 391
\re			       
Yokogawa J., Imanishi K., Tsujimoto M., Nishiuchi M., Koyama K.,
Nagase F., Corbet R.H.D.\ 1999, ApJ submitted
\re			       
Yokogawa J., Koyama K.\ 1998, IAU Circ.\ 6853

\clearpage
\begin{table*}[t]
 \begin{center}
  Table~1.\hspace{4pt}Best-fit parameters of the spectrum.\\
 \end{center}
 \vspace{6pt}
 \begin{tabular*}{\textwidth}{@{\hspace{\tabcolsep}
  \extracolsep{\fill}}p{6pc}ccc}
  \hline\hline\\ [-6pt]
  Detector  & Photon index$^{\dagger}$ & Absorption$^{\dagger}$ &
  Reduced $\chi^{2}$(d.o.f.)\\ 
  & $\Gamma$ & $N_{\rm H}(10^{21}$~cm$^{-2})$ & \\
  [4pt]\hline\\[-6pt]
  GIS \dotfill & $0.86 \pm 0.23$ & $4.8 \pm 3.7$ & 0.60 (26) \\
  SIS \dotfill & $0.61 \pm 0.22$ & $2.6 \pm 2.4$ & 1.08 (14) \\
  GIS$+$SIS \dotfill & $0.75 \pm 0.19$ & $3.6 \pm 1.9$ & 0.72 (40) \\
  \hline
 \end{tabular*}
 \vspace{6pt}\par\noindent
 $\dagger$ Errors are for 90\% confidence level.\\
\end{table*}
\begin{table*}[t]
 \begin{center}
  Table~2.\hspace{4pt}Flux history of AX J0058$-$7203.\\
 \end{center}
 \vspace{6pt}
 \begin{tabular*}{\textwidth}{@{\hspace{\tabcolsep}
  \extracolsep{\fill}}p{6pc}ccrr}
  \hline\hline\\ [-6pt]
  Satellite & Detector & Date & \multicolumn{1}{c}{Count
  rate$^{\dagger}$} & \multicolumn{1}{c}{Predicted X-ray
  flux$^{\dagger}$~$^{\ddagger}$} \\
  & & & \multicolumn{1}{c}{cnts~s$^{-1}$} &
  \multicolumn{1}{c}{erg~cm$^{-2}$~s$^{-1}$} \\
  [4pt]\hline\\[-6pt]
  Einstein & IPC  & 1979/11/13 & (1.0 $\pm$ 0.1) $\times 10^{-4}$ &
  (5.4 $\pm$ 0.5) $\times 10^{-14}$\\
           &      & 1980/03/15 & $<$ 1.9 $\times 10^{-4}$ & 
  $<$ 1.0 $\times 10^{-13}$\\ 
  ROSAT    & PSPC & 1991/10/08 & (3.3 $\pm$ 0.9) $\times 10^{-3}$ &
  (6.4 $\pm$ 1.8) $\times 10^{-13}$ \\
           &      & 1992/04/17 & $<$ 1.7 $\times 10^{-3}$ & 
  $<$ 3.3 $\times 10^{-13}$ \\
           &      & 1993/03/29 & $<$ 1.2 $\times 10^{-4}$ & 
  $<$ 2.4 $\times 10^{-14}$ \\
           &      & 1993/05/12 & (8.4 $\pm$ 11.5) $\times 10^{-4}$ &     
  (1.6 $\pm$ 2.2) $\times 10^{-13}$ \\
           &      & 1993/10/01 & (1.9 $\pm$ 1.0) $\times 10^{-3}$ &
  (3.7 $\pm$ 2.0) $\times 10^{-13}$ \\
           &      & 1994/05/05 & (2.8 $\pm$ 1.5) $\times 10^{-3}$ &
  (5.4 $\pm$ 2.9) $\times 10^{-13}$ \\
           & HRI  & 1994/04/18 & $<$ 5.7 $\times 10^{-3}$ & 
  $<$ 3.1 $\times 10^{-12}$ \\
           &      & 1994/04/19 & $<$ 1.7 $\times 10^{-3}$ & 
  $<$ 9.2 $\times 10^{-13}$ \\
           &      & 1994/10/05 & $<$ 4.9 $\times 10^{-3}$ & 
  $<$ 2.6 $\times 10^{-12}$ \\
           &      & 1995/04/12 & (2.6 $\pm$ 2.9) $\times 10^{-3}$ &
  (1.4 $\pm$ 1.6) $\times 10^{-12}$ \\
           &      & 1995/04/13 & $<$ 8.5 $\times 10^{-4}$ & 
  $<$ 4.6 $\times 10^{-13}$ \\
           &      & 1995/04/14 & $<$ 1.0 $\times 10^{-3}$ & 
  $<$ 5.4 $\times 10^{-13}$ \\
  \hline
  ASCA     & GIS  & 1997/11/14 & 1.4 $\times 10^{-2}$ & 3.2 $\times 10^{-12}$\\
  \hline
 \end{tabular*}
 \vspace{6pt}\par\noindent
 $\dagger$ Count rate with 1-$\sigma$ error in a circle of
 $1^{\prime}$ radius around the source position, after subtracting
 the background count rate taken in a neighboring off-source
 region. Upper limits for 1-$\sigma$ significance level are given
 for observations in which the source was not detected.\\
 \par\noindent
 $\ddagger$  X-ray flux 1-$\sigma$ error in the 0.7--10.0 keV is
 simulated from the count rate by the {\tt pimms} software,
 assuming no change from the ASCA spectrum. For those not detected,
 the upper limit is given. The ASCA result, together with the count
 rate, is also shown at the last row.\\
\end{table*}

\end{document}